%% file: 0main.tex
\documentclass[runningheads]{llncs}
\usepackage[T1]{fontenc}

\usepackage{hyperref}
\usepackage{graphicx}
\usepackage{amsmath} 
\usepackage{booktabs}
\usepackage{amsmath}
\usepackage{amssymb}
\usepackage{pifont}
\usepackage{xcolor}
\usepackage{graphicx}
\usepackage{subcaption}
\usepackage{float}
\usepackage{listings}
\usepackage{caption}
\lstdefinestyle{ircompact}{
  basicstyle=\ttfamily\tiny,
  columns=fullflexible,
  keepspaces=true,
  breaklines=true,
  breakatwhitespace=false,
  frame=single,
  framerule=0.3pt,
  xleftmargin=1pt,
  xrightmargin=1pt,
  aboveskip=2pt,
  belowskip=2pt,
  numbers=none,
  showstringspaces=false
}

\begin{document}
\title{Confidence-Guided Protocol IR for LLM-Aided Security Protocol Modeling}
%
%
\author{Siqi Li\inst{1,2}\orcidID{0009-0001-9896-243X} \and
Yufan Cai\inst{1}\orcidID{0009-0008-7579-0824} \and
Hongshu Wang\inst{1}\orcidID{0009-0006-0198-148X} \and
Xinyue Zuo\inst{1}\orcidID{0009-0008-4411-3054}\and
Zhe Hou\inst{3}\orcidID{0000-0001-7164-0580}\and
Jin Song Dong\inst{1}\orcidID{0000-0002-6512-8326}
}
\authorrunning{S. Li et al.}
%
\institute{National University of Singapore, Singapore \and
Beijing Normal-Hong Kong Baptist University, China \and
Griffith University\\
\email{t330013016@mail.bnbu.edu.cn}\\
\email{\{caiyf, hongshu.wang, zuoxy, dcsdjs\}@nus.edu.sg}\\
\email{z.hou@griffith.edu.au}\\
}
\maketitle              

\begin{abstract}
Large language models offer a promising interface for translating natural-language protocol descriptions into formal security models, but their outputs remain difficult to trust without expert validation.
In this paper, we present a human-in-the-loop framework for generating Tamarin-verifiable formal models of security protocols. 
Our key observation is that the main correctness bottleneck is the semantic accuracy rather than the syntactic validity of the intermediate protocol representation. 
To address this problem, we introduce a protocol \textbf{intermediate representation (IR)} that serves as a human-auditable semantic checkpoint between natural-language parsing and formal model generation. 
The IR explicitly captures protocol participants, message flows, value provenance, cryptographic operations, proof targets, and compromise assumptions. 
We further design an interactive interface that highlights uncertain fields and guides users to inspect the most critical semantic decisions based on model confidence before model generation. 
Rather than replacing formal-methods experts, our approach uses LLMs to produce auditable semantic drafts while leveraging verification tools to check the resulting formal models. 

\end{abstract}

\input{1intro}
\input{3.1approach}
\input{4experiment}
\input{2case}

\input{5discussion}

\input{7conclusion}


%
%
%
\bibliographystyle{splncs04}
\bibliography{reference}
%




\end{document}

%% file: 1intro.tex
\section{Introduction}
Security protocols are communication workflows that allow distributed parties to exchange information securely over untrusted networks. 
They are the foundation of modern digital systems, including online finance, e-commerce, cloud services, blockchains, and decentralized applications. 
Since these protocols are deployed in adversarial environments, even small design flaws may lead to severe security breaches and financial losses. 
A notable example is the 2016 DAO attack on Ethereum, where a re-entrancy vulnerability in a smart contract resulted in a loss of approximately \$60 million~\cite{praitheeshan2019survey}. Consequently, informal reasoning and testing are often insufficient for establishing deployable security guarantees, especially for blockchain protocols and smart contracts that are difficult or impossible to patch after deployment. 
Formal verification is therefore desirable because it provides a rigorous way to analyze protocol behavior and establish correctness and security guarantees~\cite{yuiceccs2025}.
Formal verification tools such as Tamarin~\cite{meier2013tamarin}, ProVerif~\cite{blanchet2018proverif}, and SAPIC+~\cite{cheval2022sapic+} provide strong support for symbolic protocol analysis. 
They can verify secrecy, authentication, and correspondence properties, and they can generate counterexamples when these properties do not hold. 
However, the reliability of these results depends critically on the correctness of the underlying formal model, which is typically constructed manually by a human expert. 
Automatically constructing such models from informal protocol requirements, including natural-language descriptions, message diagrams, standards documents, and research papers, remains challenging. 
A model may be syntactically valid and verified, while still encoding the wrong protocol semantics.

Recent work has explored the integration of large language models with formal methods for tasks such as formal specification development, model construction, verification-guided repair, and trustworthy agent development~\cite{fuggitti2023nl2ltl,zuo2025pat,zhang2025position,wang2026event,chen2026modelwisdom}. 
Motivated by this broader trend, LLMs offer a promising way to reduce the burden of security protocol modeling~\cite{mao2025llm}. 
Given a natural-language protocol description, an LLM can generate structured summaries, infer message flows, and even produce candidate SAPIC+ models. 
However, directly generating Tamarin code from natural language remains unreliable. 
The core difficulty is not merely syntactic generation, but semantic faithfulness: the generated model may compile and verify while still misrepresenting the intended protocol.
Once the relevant protocol semantics are correctly captured, generating a formal model can be made largely systematic. 
Fresh values can be translated into declarations, long-term state into setup processes, messages into input/output actions, checks into conditional tests, events into symbolic annotations, and proof targets into lemmas. 
The difficult part is ensuring that these objects faithfully represent the intended protocol before they are translated into a formal model.

To address this challenge, we propose a confidence-guided, human-auditable Protocol Intermediate Representation, called Protocol IR, for LLM-aided modeling of security protocols. 
Instead of treating the LLM output as the final formal model, our framework treats it as a structured semantic draft. The Protocol IR records the modeling decisions that determine the meaning of the generated formal model, including value provenance, initial knowledge, message structure, cryptographic checks, event placement, proof intent, compromise assumptions, and expected counterexamples. 
In this way, the IR serves as a semantic checkpoint between informal protocol descriptions and the generation of trustworthy formal models.
The proposed framework aims to reduce the effort and expertise required for formal modeling while still allowing users to intervene in the modeling process. 
An interface displays the auditable semantic draft together with confidence information for its components. 
Our framework makes this trust boundary explicit: the initial IR is untrusted, but the reviewed IR becomes the trusted source for model generation.
To reduce the cost of human review, we introduce confidence-guided inspection. 
Each IR field is annotated with confidence information based on evidence support, consistency with other IR entries, and semantic risk. 
For example, a field directly supported by a source sentence may receive high confidence, while a value used before generation, an event placed before a check, or a receiver assumed to know an underivable term should be highlighted as risky. 
Since not all fields are equally important, low confidence in proof targets, checks, events, or compromise assumptions is prioritized over low confidence in descriptive fields. The interface uses these signals to guide users toward the IR fields most likely to affect verification correctness.

In summary, this paper makes the following contributions:
\begin{itemize}
\item We propose a  confidence-guided Protocol IR that exposes security-critical modeling decisions, including value provenance, initial knowledge, message structure, cryptographic checks, event placement, proof intent, compromise assumptions, and expected counterexamples.
\item We implement an end-to-end framework that translates natural-language protocol descriptions into Protocol IR, supports interactive human review through a user interface, automatically generates formal models from the reviewed IR, and invokes backend verification and repair.
\item We evaluate the framework on representative security protocol modeling tasks and non-standard cases, studying the correctness of generated IR fields, the effectiveness of confidence-guided review, and the quality of the resulting formal models and verification outcomes.
\end{itemize}

The implementation is available on GitHub.\footnote{\url{https://github.com/laplace1002/TamarinAgent.git}}

%% file: 3.1approach.tex
\section{Approach}
\label{sec:approach}

\begin{figure}[t]
    \centering
    \includegraphics[width=1\linewidth]{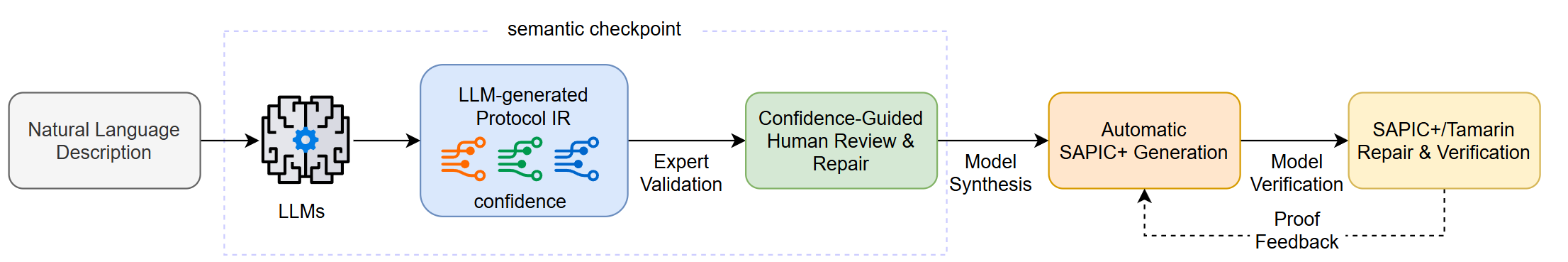}
    \caption{Overview of the proposed workflow. The Protocol IR serves as a semantic checkpoint
    between unreliable natural-language interpretation and formal model generation.}
    \label{fig:workflow}
\end{figure}

Figure~\ref{fig:workflow} shows the overall workflow of our approach.
Given a natural-language protocol description, the system first invokes an LLM to extract
protocol semantics into a structured Protocol IR. The IR captures security-critical modeling
decisions, including fresh values, long-term state, message structure, checks, event placement,
proof targets, and compromise assumptions. Each generated IR field is associated with provenance
information and confidence signals, which are later used to guide human review.
The user then inspects the generated IR through a review interface. Instead of requiring the user
to read low-level SAPIC+ code line by line, the interface exposes high-level protocol concepts and
highlights uncertain or semantically risky fields. After the user confirms or edits the IR, the
reviewed IR becomes the trusted semantic source for automatic SAPIC+ generation. The generated
model is then checked by Tamarin. When syntactic or backend-specific errors are detected, the
system attempts repair while preserving the reviewed protocol semantics.

\subsection{Protocol IR}
\label{subsec:protocol-ir}

The Protocol IR operationalizes the semantic checkpoint in our workflow. It organizes protocol
semantics into structured components that are both reviewable by humans and usable for downstream
SAPIC+ generation. Unlike raw natural-language descriptions, the IR makes modeling choices
explicit: which values are fresh, which values belong to long-term setup, how messages are
constructed, which checks are performed, where security events are placed, what properties should be proved, and what compromise assumptions are allowed.

\paragraph{Fresh values.}
The IR records values generated during protocol execution, such as nonces, random challenges,
ephemeral keys, and session keys. For each fresh value, the IR records its symbolic name, owner,
and purpose. The symbolic name is used later in message terms and proof events. The owner records
which role generates the value, while the purpose explains why the value exists, such as freshness,
challenge generation, or session-key establishment. A key distinction is whether a value is
generated freshly during protocol execution or belongs to setup or long-term state. Confusing
these two cases can substantially change the security meaning of the generated model.

\paragraph{Cryptographic declarations.}
The IR records the cryptographic vocabulary required by the protocol, including built-in theories
such as asymmetric encryption, hashing, or symmetric encryption, as well as protocol-specific
functions, equations, and assumptions when needed. These declarations determine which symbolic
operations are available when generating SAPIC+ code.

\paragraph{Long-term state and setup assumptions.}
The IR records long-term secrets, public keys, shared keys, role identities, and setup-generated
state. For each value, it specifies which role owns it, whether it has a corresponding public term,
and whether it may be compromised. This component is essential for constructing a faithful attacker
model, since changing a setup assumption may alter whether an attack is possible.

\paragraph{Messages.}
The IR describes each protocol message using a structured representation of sender, receiver,
message fields, protection mode, and symbolic term. The protection mode records whether the
message is public, encrypted, signed, authenticated by a MAC, or unknown. The symbolic term is used
by downstream SAPIC+ generation, while the natural-language meaning records the intended
interpretation of the message. The IR also stores derived provenance information, such as which
values the sender must know to construct the message and which values the receiver is expected to derive after parsing it. 

\paragraph{Actions.}
The IR contains role-local protocol transitions that map message-level descriptions to role
processes.
An action records the executing role, generated values, incoming and outgoing messages,
checks performed by the role, and events emitted after the transition. 
These actions provide the bridge between protocol-level descriptions and the process structure later emitted as code.

\paragraph{Checks and Events.}
The IR explicitly represents verification operations, including equality checks, hash comparisons, MAC verification, signature verification, decryption success, and reconstruction checks. Each check records the role that performs it, the condition being checked, the source message providing the relevant evidence, and the associated role-local action. Checks are security-critical because symbolic events should usually be placed only after the relevant checks have succeeded.
The IR records symbolic events used by security lemmas, such as \texttt{Running}, \texttt{Commit}, \texttt{Accept}, \texttt{Secret}, and \texttt{Reveal}. Each event records its name, role, arguments, and intended placement. Together with surrounding actions and checks, these fields make event placement auditable. For example, an authentication event should not be triggered immediately after receiving a message if the role has not yet verified its contents.

\paragraph{Proof targets.}
The IR records the intended verification goals, including lemma names, goal types, trace kinds,
related events, secrecy targets, authentication targets, expected verification results, and
expected counterexamples. 
Proof targets are among the most important fields for human review.

\paragraph{Compromise assumptions and attack surface.}
The IR records attacker capabilities, reveal rules, compromise assumptions, and expected attacks.
This prevents the repair process from accidentally removing an intended attack or weakening the
Dolev--Yao adversary simply to make a proof succeed.

\begin{figure}[!t]
\centering
\begin{center}
\begin{minipage}[t]{0.49\linewidth}
\vspace{0pt}
\begin{lstlisting}[style=ircompact]
{
 "schema":"protocol_ir_v1",
 "protocol_name":"Example",
 "roles":["C","S"],
 "crypto":{
  "builtins":[
   "asymmetric-encryption",
   "hashing",
   "symmetric-encryption"
  ],
  "functions":[],
  "equations":[],
  "assumptions":[
   "Use the standard Dolev-Yao adversarial network model."
  ]
 },
 "fresh_terms":[
  {"name":"~k","owner":"C",
   "purpose":"fresh symmetric session key"}
 ],
 "long_term_keys":[
  {"name":"ltkS","owner":"S",
   "public_term":"pk(ltkS)",
   "policy":"server private key is not revealed"}
 ],
 ...
\end{lstlisting}
\end{minipage}
\hfill
\begin{minipage}[t]{0.49\linewidth}
\vspace{0pt}
\begin{lstlisting}[style=ircompact]
 "messages":[
  {
   "label":"M1",
   "from":"C",
   "to":"S",
   "protection":"asymmetric-encryption",
   "term":"aenc(~k, pk(ltkS))",
   "meaning":"C sends a fresh key encrypted for S",
   "_hidden_or_derived":{
    "sender_knows":["~k","pk(ltkS)"],
    "receiver_can_decrypt":true
   }
  }
 ],
 ...
 "field_evidence":[
  {
   "field_path":"messages.0.protection",
   "source_quote":"encrypts it with the public key pkS",
   "evidence_kind":"direct",
   "priority_llm":0.3,
   "evidence_confidence_score":1.0,
   "consistency_confidence_score":1.0,
   "semantic_impact_score":0.9
  }
 ]
}
\end{lstlisting}
\end{minipage}
\captionof{lstlisting}{Partial raw Protocol IR for the ``Example'' protocol. The left column shows protocol-level declarations and setup information, while the right column shows message-level fields and LLM-provided field evidence. Ellipses indicate omitted IR sections such as checks, events, actions, proof targets, compromise assumptions, and semantic constraints.}
\label{lst:raw-ir-example}
\end{center}
\end{figure}

\paragraph{Review Interface}
\label{subsec:review-interface}

The review interface presents the reviewable IR as editable structured sections. Its purpose is
not to expose all low-level model-generation details, but to help the user inspect the semantic
choices that determine the correctness of the final SAPIC+ model. The same IR components
introduced in \S\ref{subsec:protocol-ir} are rendered as review pages, including fresh values,
setup and long-term state, messages, checks, events, proof targets, compromise assumptions, and
expected attack surfaces.
The left sidebar shows the current protocol name, a color legend for field review status, and
navigation groups for the workflow. These groups include a start page for natural-language input,
review pages for editable IR sections, and generation pages for SAPIC+ output and Tamarin results.
Each navigation item may display a workflow status badge, such as current, pending, or done, as
well as the number of unresolved review cells.

\begin{figure}[t]
    \centering
    \includegraphics[width=\linewidth]{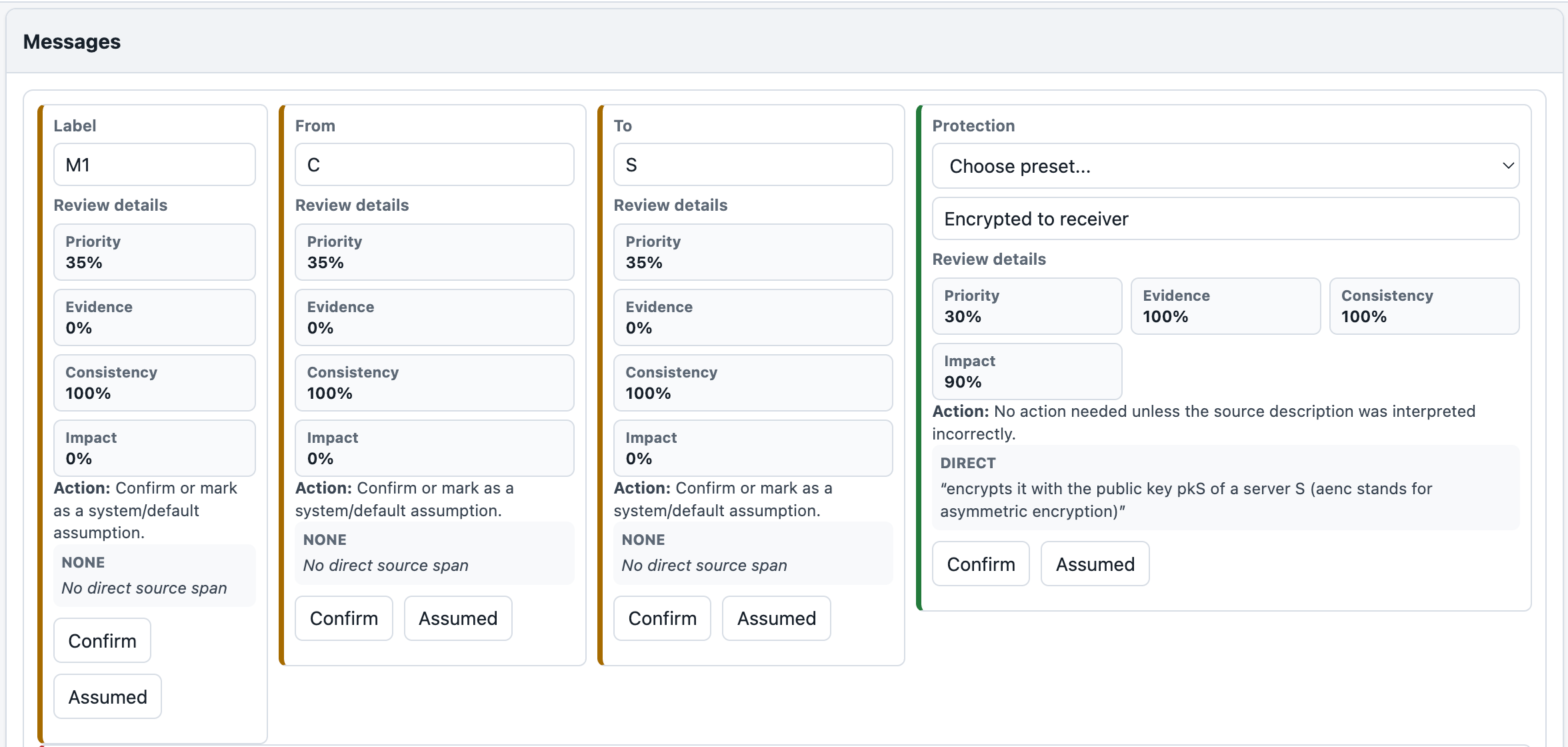}
    \caption{\textbf{Message header and protection cell in the review interface.}
    The message page exposes editable cells for the reviewable message fields. Clicking
    ``Review details'' expands field-level diagnostic metadata used for confidence-guided review,
    including confidence scores, review priority, recommended action, and source evidence.}
    \label{fig:review-ui-up}
\end{figure}

\begin{figure}[t]
    \centering
    \includegraphics[width=\linewidth]{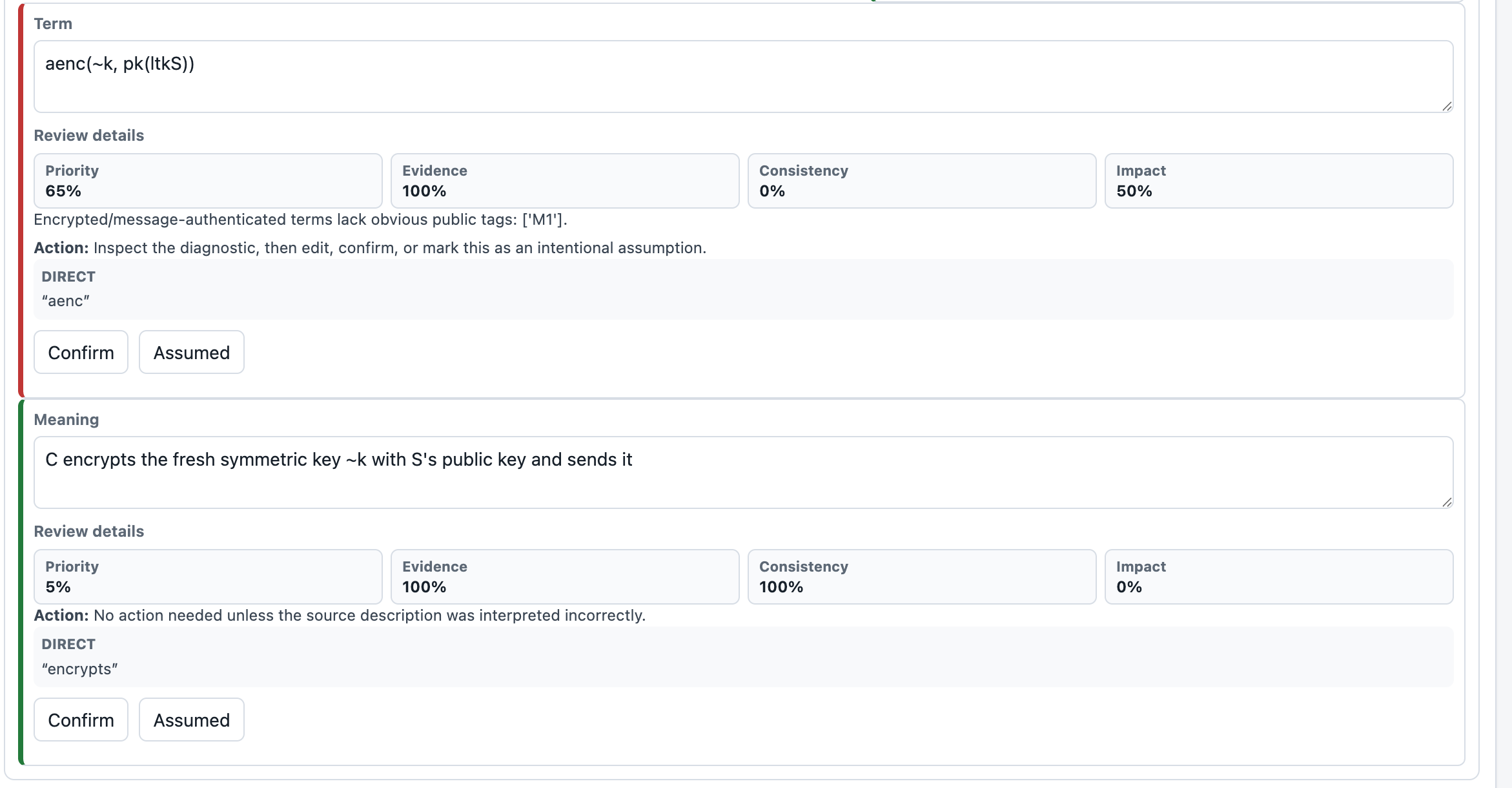}
    \caption{\textbf{Message term and meaning review details.}
    This view continues the review of message \texttt{M1}, where role \texttt{C} sends the fresh
    key \texttt{\string~k} to role \texttt{S}. The term cell contains the symbolic message used
    by SAPIC+ generation, while the meaning cell records the corresponding natural-language
    interpretation.}
    \label{fig:review-ui-down}
\end{figure}

\paragraph{Natural-language input.}
The first interface page allows the user to create a Protocol IR directly from a natural-language
description. It contains fields for the protocol name, difficulty label, protocol description,
assumptions, and verification goals. The protocol description is the only required input. Optional
assumptions can specify attacker-model or setup information, such as Dolev--Yao network control
or trusted public-key setup. Optional goals can specify target lemma names, goal types, trace
kinds, and expected results. If goals are omitted, the planner infers candidate proof targets from
the protocol description.

After the user presses ``Generate IR / Contract'', the system invokes the planner to produce a
Protocol IR candidate. It then validates the IR, derives field-level review metadata, and
constructs the reviewable modeling contract. The validation pass checks for semantic risks such
as missing required proof events, events emitted before the corresponding check or decryption,
conflicting value classes, and derivability problems. These diagnostics can force a field into
``needs review'' even when the LLM-provided confidence is high.

\paragraph{Prepared workflow and IR review pages.}
This page is used in the experimental workflow to load an already prepared protocol case from the
local workflow library. It allows the same review and generation pipeline to be applied
consistently across benchmark protocols.
The review pages render the editable cells of each IR component. For example, the message page
exposes the reviewable message fields defined in \S\ref{subsec:protocol-ir}, as shown in
Figures~\ref{fig:review-ui-up} and~\ref{fig:review-ui-down}. The user can confirm a generated
field, edit it, or mark it as an intentional modeling assumption.

\paragraph{Per-cell review details.}
Every visible editable cell contains a ``Review details'' panel. This panel displays the
field-level confidence signals, review priority, validation diagnostics, recommended reviewer
action, and source evidence when available. If no direct source span is found, the panel explicitly
reports that the field is inferred or assumed. The ``Confirm'' button marks the cell as manually
validated, while the ``Assumed'' button records that the user intentionally accepts the value as a
modeling assumption.

\paragraph{SAPIC+ generation and Tamarin results.}
After review, the user saves the edited modeling contract. The system then materializes the
reviewed contract back into a reviewed Protocol IR and uses it for SAPIC+ generation. The SAPIC+
page provides controls to generate the model, run a repair-and-verify loop, and invoke Tamarin
proofs. The Tamarin results page displays compile status, proof status, warnings, mismatched
expected results, and the generated Tamarin code. These outputs allow the user to check whether a
Tamarin-clean model has been generated and whether the expected proof outcomes are obtained.

\subsection{LLM-Based IR Construction}
\label{subsec:llm-ir-construction}

Given a natural-language protocol description, the LLM is prompted to generate a Protocol IR
rather than SAPIC+ code directly. This design deliberately separates semantic interpretation from
formal code generation. The LLM is responsible for extracting a structured candidate
interpretation of the protocol, including roles, values, messages, checks, events, proof targets,
and compromise assumptions.
For each generated IR field, the system asks the LLM to provide supporting evidence from the
source description when possible. For example, if the LLM claims that a nonce is freshly generated
by the initiator, it should identify the corresponding textual evidence. If the field is inferred
rather than explicitly stated, the IR marks the field as inferred. If no direct evidence exists,
the field is marked as an assumption.
This evidence-aware construction makes the IR auditable. The user does not need to treat the LLM
output as an opaque answer. Instead, the user can inspect both the generated field and the reason
why the system believes the field is correct.

\subsection{Confidence-Guided Human Review}
\label{subsec:confidence-review}

A complete Protocol IR may contain many fields. Exhaustively reviewing every field can be
expensive, especially for complex protocols. To reduce review effort, our framework assigns
confidence and review priority to IR fields. The review priority is derived from three kinds of
signals.

\paragraph{Evidence confidence.}
Evidence confidence measures whether a field is directly supported by the source description. A
field receives high evidence confidence if it is supported by an explicit source span, medium
confidence if it is inferred from nearby context, and low confidence if it has no direct textual
support.

\paragraph{Consistency confidence.}
Consistency confidence measures whether a field is compatible with other IR entries. Examples of
low consistency include a value used before it is generated, a role verifying a term it cannot
derive, an event placed before the corresponding check, or a value classified both as a
per-session nonce and a long-term secret.

\paragraph{Semantic impact.}
Semantic impact measures how much an error in a field may affect the meaning of the final
verification result. Errors in proof targets, event placement, checks, or compromise assumptions
can make the final verification result meaningless. Therefore, even a moderately uncertain field
should receive high review priority if its semantic impact is large.

\paragraph{Review priority.}
Conceptually, the review priority of a field is computed as
$\mathsf{priority}(f)
=
\max(
1-\mathsf{evidence}(f),
1-\mathsf{consistency}(f)
)
\times
\mathsf{impact}(f)
$.
This score is used to guide attention.
Fields with high priority are highlighted in the interface and should be inspected first.

\subsection{Formal Model Generation from Reviewed IR}
\label{subsec:sapic-generation}

After human review, the confirmed IR is used as the semantic source for SAPIC+ generation. Because
the IR is structured around protocol concepts, the translation can be performed systematically.
Fresh values are translated into \texttt{new} declarations. Long-term state and setup assumptions
are translated into setup processes, persistent facts, or role parameters. Message entries are
translated into input and output actions. Cryptographic checks are translated into \texttt{let},
equality, or conditional checks. Events are translated into SAPIC+ event annotations. Proof
targets are translated into Tamarin lemmas. Compromise assumptions are translated into reveal
rules and attacker-knowledge declarations.
For medium and hard cases, the user can optionally enable abstraction hints as proof-engineering
assistance. Once enabled, the backend retrieves matching hints from a predefined hint library using
the reviewed IR and proof targets. These hints describe modeling patterns such as compact event
payloads, bounded role topology, transcript summaries, and source lemma shapes.
The generation prompt is designed to preserve the reviewed IR semantics. In particular, it should
not introduce additional checks, remove attacker capabilities, or change event placement unless
such changes are reflected in the IR and confirmed by the user.

\subsection{Repair and Verification}
\label{subsec:repair-verification}

The generated SAPIC+ model is submitted to Tamarin for parsing and verification. Some generated
models may require repair due to syntactic errors, backend-specific restrictions, Tamarin
well-formedness warnings, or proof-lint issues. Our framework allows automatic repair at this
stage, but restricts the repair scope to preserve the reviewed semantics.
We distinguish two kinds of repair. Syntactic repair fixes malformed SAPIC+ constructs, incorrect
declarations, backend compatibility issues, or translation-level inconsistencies in events,
checks, and lemmas when the fix is supported by the reviewed IR. Semantic repair changes the
protocol meaning, such as moving events, adding checks, changing attacker capabilities, or
modifying proof targets. Syntactic repair can be automated safely because it does not alter the
intended protocol semantics. Semantic repair requires returning to the IR and asking the user to
review the affected field.
This distinction prevents the system from optimizing for proof success at the cost of model faithfulness. 
If Tamarin finds a counterexample, the system does not automatically treat it as a
modeling bug. Instead, the counterexample is compared against the expected attacks and compromise assumptions recorded in the IR. 

%% file: 4experiment.tex
\section{Evaluation}
\label{sec:evaluation}

In this section, we evaluate whether our framework can help users
construct trustworthy formal protocol models from natural-language descriptions. 
Following the evaluation structure of prior work on
LLM-aided automatic symbolic modeling~\cite{mao2025llm}, we evaluate
our approach at multiple levels: the quality of the generated protocol
IR, the effectiveness of confidence-guided human review, and the correctness of the generated SAPIC+ models.

Our evaluation is designed to answer the following research questions:
\begin{description}
    \item[RQ1: IR Extraction Quality.]
    How accurately can the LLM extract a protocol IR from a natural-language protocol description?

    \item[RQ2: End-to-End Model Generation.]
    Given a reviewed protocol IR, can our framework generate SAPIC+ models that are accepted by Tamarin?

    \item[RQ3: Generated versus Manual Models.]
    How do generated models differ from manually developed Tamarin models and what do these differences reveal about the limits of LLM-assisted modeling?

    \item[RQ4: Cost Analysis.]
    What is the cost of introducing the Protocol IR and confidence-guided review into the modeling workflow?
    
\end{description}

\paragraph{Implementation.}
We implemented our framework as an interactive modeling assistant.
All experiments are conducted on a MacBook Pro equipped with an Apple M5 Pro chip, a 15-core CPU, a 16-core integrated GPU, 24 GB of RAM, and macOS 26.5 (Build 25F71). For LLM calls, we use provider-hosted chat-completion APIs as served during June 7--8, 2026. 
We evaluate our system using the following LLMs: DeepSeek V4 Pro~\cite{deepseekai2026deepseekv4}, GPT-4o~\cite{hurst2024gpt}, GPT-5.5~\cite{hurst2024gpt}, and Llama-3.3-70b-instruct~\cite{grattafiori2024llama}.
All generated SAPIC+ models are checked using Tamarin version\texttt{1.12.0} with Maude version \texttt{3.5.1}. 

\paragraph{Metrics}
\label{subsec:metrics}

We follow prior work's exact-coverage, boundedness-check, and property-success evaluation style~\cite{mao2025llm}, but adapt the counting unit to our UI-visible IR review cells.
For RQ1, Table~\ref{tab:ir-quality} reports exact-match rate (EMR), semantic accuracy (SA), well-formedness rate (WFR), and semantic repair count ($\#\epsilon$). EMR adapts exact coverage to UI-visible IR cells:
\(\mathrm{EMR}=N_{\mathrm{match}}/N_{\mathrm{cell}}\), where
$N_{\mathrm{cell}}$ is the number of evaluated UI-visible cells. SA is also
cell-level and treats exact and semantically acceptable labeled cells as
positive:
\(\mathrm{SA}=(N_{\mathrm{correct}}+N_{\mathrm{acceptable}})/N_{\mathrm{labeled\text{-}cell}}\),
where $N_{\mathrm{labeled\text{-}cell}}$ excludes unlabeled or invalid rows.
WFR instead performs a role-local provenance check over symbolic-value uses:
\(\mathrm{WFR}=N_{\mathrm{provenance\text{-}ok}}/N_{\mathrm{provenance}}\).
It therefore has a different denominator and measures neither global semantic
correctness nor SAPIC+ syntactic acceptance. Finally,
\(\#\epsilon=N_{\mathrm{labeled\text{-}cell}}-N_{\mathrm{correct}}-N_{\mathrm{acceptable}}\)
counts existing UI-visible cells that require semantic repair.
Because the number of emitted cells and provenance uses depends on the
generated IR, these RQ1 measures are output-conditioned diagnostics rather
than recall over a fixed set of reference obligations.
For RQ2, Table~\ref{tab:e2e} summarizes SAPIC+ acceptance and verification
success as case-level ratios in the totals.
For RQ3, we compare GPT-5.5-generated models with manual models using the structural alignment metrics defined below. For this comparison, comments and lemmas are removed, identifiers and constants are alpha-normalized, and equivalent cryptographic aliases are normalized. We apply these metrics to four dimensions corresponding to explicit modeling decisions, including value provenance and setup, message payload structure, cryptographic dependencies, and verification checks. Each dimension is represented as a multiset of structural atoms. For generated atoms $G$ and reference atoms $R$, the matched atoms are $M = G \cap R$. We report
$\mathrm{Precision} = |M|/|G|$, $\mathrm{Recall} = |M|/|R|$, $
\mathrm{F1} = 2PR/(P+R)$. Precision penalizes unsupported additions, while recall captures details omitted from the generated model.
For RQ4, we report UI-visible manual edit counts
and runtime overhead for SAPIC+ generation and Tamarin verification.

\subsection{RQ1: IR Extraction Quality}
\label{subsec:rq1}

For each protocol description, we prompt the LLM to generate a protocol IR.
We compare the generated IR against the manually reviewed reference IR. We
evaluate both exact field-level matching and semantic correctness. Unlike
direct SAPIC+ generation, the IR is designed to expose semantic information
that is important for formal modeling, including value provenance,
verification targets, role-local knowledge, and compromise assumptions.
  
Table~\ref{tab:ir-quality} reports the IR extraction results.
We observe that exact match for each IR field remains challenging because the average EMR is mostly under 55.0\%, even for strong models. However, semantic accuracy is higher, especially for GPT-5.5, indicating that many emitted cells that do not exactly match the reference are still semantically acceptable. This does not account for reference details that the model fails to emit. GPT-4o achieves the best average EMR (51.9\%) and the fewest $\epsilon$ among the non-GPT-5.5 models. DeepSeek V4 Pro has slightly higher semantic accuracy than GPT-4o and Llama, but requires more missing-cell corrections. GPT-5.5 achieves near-perfect semantic accuracy on the annotated UI cells, suggesting that it often captures protocol meaning. Its lower average WFR (44.8\%) captures a different issue. WFR checks whether every role-local value use has an explicit source. For example, a cell may correctly state that a server computes its response from a nonce $n$ and therefore receive an acceptable SA label. If the server role does not explicitly obtain $n$ through an \texttt{in} action
or carry it in local state, the corresponding use fails the WFR check.

\begin{table*}[t]
\centering
\caption{IR extraction quality at the UI-visible cell level.}
\label{tab:ir-quality}
  \tiny
  \setlength{\tabcolsep}{1pt}
  \begin{tabular}{lrrrrrrrrrrrrrrrr}
  \toprule
   & \multicolumn{4}{c}{\textbf{DeepSeek V4 Pro}} & \multicolumn{4}{c}{\textbf{GPT-4o}} & \multicolumn{4}{c}{\textbf{Llama-3.3-70b-instruct}} & \multicolumn{4}{c}{\textbf{GPT-5.5}} \\
  \cmidrule(lr){2-5}\cmidrule(lr){6-9}\cmidrule(lr){10-13}\cmidrule(lr){14-17}
  \textbf{Protocol} & \textbf{EMR} & \textbf{SA} & \textbf{WFR} & \textbf{\#$\epsilon$} & \textbf{EMR} & \textbf{SA} & \textbf{WFR} & \textbf{\#$\epsilon$} & \textbf{EMR} & \textbf{SA} & \textbf{WFR} & \textbf{\#$\epsilon$} & \textbf{EMR} & \textbf{SA} & \textbf{WFR} & \textbf{\#$\epsilon$} \\
  \midrule
Example & 50.6\% & 94.8\% & 100.0\% & 4 & 62.3\% & 93.4\% & 100.0\% & 4 & 60.7\% & 93.4\% & 100.0\% & 4 & 52.0\% & 100.0\% & 52.9\% & 0 \\
NSPK & 52.3\% & 80.2\% & 85.7\% & 17 & 56.0\% & 75.0\% & 100.0\% & 21 & 54.2\% & 75.0\% & 76.9\% & 18 & 35.9\% & 100.0\% & 65.7\% & 0 \\
Naxos & 32.6\% & 85.7\% & 55.6\% & 7 & 49.0\% & 89.8\% & 100.0\% & 5 & 47.5\% & 74.6\% & 100.0\% & 15 & 44.2\% & 100.0\% & 44.4\% & 0 \\
Toy & 35.3\% & 95.3\% & 95.8\% & 4 & 49.0\% & 82.3\% & 100.0\% & 9 & 43.7\% & 76.1\% & 100.0\% & 17 & 30.4\% & 100.0\% & 55.6\% & 0 \\
Woo and Lam & 53.4\% & 84.1\% & 100.0\% & 14 & 55.1\% & 73.1\% & 89.5\% & 21 & 57.7\% & 73.1\% & 88.2\% & 21 & 38.9\% & 100.0\% & 51.1\% & 0 \\
Sigfox & 44.6\% & 78.5\% & 83.3\% & 14 & 47.3\% & 90.9\% & 88.9\% & 5 & 43.9\% & 70.2\% & 100.0\% & 17 & 27.4\% & 100.0\% & 82.6\% & 0 \\
\midrule
X509.1 & 43.3\% & 55.7\% & 61.1\% & 43 & 52.7\% & 80.0\% & 71.4\% & 11 & 53.2\% & 79.0\% & 77.8\% & 13 & 39.0\% & 99.2\% & 34.6\% & 1 \\
Denning\-Sacco & 45.0\% & 93.6\% & 90.9\% & 7 & 53.7\% & 88.1\% & 100.0\% & 8 & 53.5\% & 85.9\% & 83.3\% & 10 & 39.5\% & 100.0\% & 95.1\% & 0 \\
Kao Chow & 63.5\% & 87.3\% & 53.8\% & 16 & 61.5\% & 79.2\% & 66.7\% & 20 & 56.2\% & 77.1\% & 68.8\% & 22 & 44.4\% & 100.0\% & 40.0\% & 0 \\
NSSK & 51.8\% & 82.9\% & 90.0\% & 28 & 67.0\% & 76.0\% & 100.0\% & 24 & 64.6\% & 79.2\% & 86.4\% & 20 & 55.3\% & 100.0\% & 35.1\% & 0 \\
Stubblebine & 45.3\% & 57.9\% & 66.7\% & 67 & 49.2\% & 63.3\% & 69.0\% & 47 & 38.8\% & 70.9\% & 70.3\% & 48 & 34.2\% & 100.0\% & 25.0\% & 0 \\
Otway Rees & 60.2\% & 93.2\% & 94.3\% & 11 & 67.0\% & 77.0\% & 66.7\% & 23 & 61.0\% & 68.0\% & 73.3\% & 32 & 46.1\% & 100.0\% & 38.5\% & 0 \\
Yahalom & 58.2\% & 81.8\% & 73.9\% & 20 & 62.2\% & 79.6\% & 68.4\% & 20 & 63.9\% & 80.2\% & 85.7\% & 17 & 40.1\% & 100.0\% & 50.9\% & 0 \\
\midrule
EDHOC & 27.9\% & 78.9\% & 73.4\% & 62 & 39.5\% & 73.7\% & 73.7\% & 30 & 44.4\% & 60.2\% & 93.3\% & 43 & 32.4\% & 100.0\% & 23.8\% & 0 \\
KEMTLS & 16.9\% & 74.3\% & 92.6\% & 87 & 34.8\% & 64.0\% & 59.1\% & 32 & 36.1\% & 54.2\% & 66.7\% & 38 & 19.1\% & 100.0\% & 32.6\% & 0 \\
LAKE & 40.5\% & 87.8\% & 41.2\% & 9 & 35.7\% & 72.9\% & 85.7\% & 19 & 40.3\% & 75.8\% & 75.0\% & 15 & 27.1\% & 100.0\% & 37.1\% & 0 \\
SPLICE & 48.2\% & 59.7\% & 85.7\% & 56 & 50.0\% & 57.6\% & 75.0\% & 39 & 47.1\% & 63.7\% & 88.2\% & 37 & 24.4\% & 100.0\% & 40.0\% & 0 \\
SSH & 37.7\% & 45.5\% & 52.9\% & 42 & 42.0\% & 76.8\% & 64.3\% & 16 & 47.4\% & 77.2\% & 75.0\% & 13 & 19.9\% & 100.0\% & 1.6\% & 0 \\
\midrule
\textbf{Average} & 44.9\% & 78.7\% & 77.6\% & 28.2 & 51.9\% & 77.4\% & 82.1\% & 19.7 & 50.8\% & 74.1\% & 83.8\% & 22.2 & 36.1\% & 99.96\% & 44.8\% & 0.1 \\
  \bottomrule
  \end{tabular}
\end{table*}

\subsection{RQ2: End-to-End SAPIC+ Model Generation}
\label{subsec:rq3}

After user inspection and repair, we translate the reviewed protocol IR into
SAPIC+. We then run Tamarin on the generated model and check the target
security properties. A case is considered successful if the generated model
is accepted by Tamarin and all target properties are verified or produce
expected attacks.
Table~\ref{tab:e2e} reports the end-to-end results. DeepSeek provides the strongest compile-success baseline, with all 18 generated models accepted by Tamarin, although only 9 of them satisfy the expected verification outcomes. In contrast, GPT-4o and Llama struggle in downstream SAPIC+ generation. GPT-5.5 achieves the highest verification success so far, with 14 out of 18 cases verified, suggesting that stronger models can improve the generation performance.

\begin{table*}[t]
\centering
\caption{End-to-end model generation and verification results for one round.}
\label{tab:e2e}
\tiny
\setlength{\tabcolsep}{1pt}
\begin{tabular}{p{1.05cm}rrrrrrrrrrrrrrrr}
\toprule
\textbf{Protocol} & \multicolumn{4}{c}{\textbf{DeepSeek}} & \multicolumn{4}{c}{\textbf{GPT-4o}} & \multicolumn{4}{c}{\textbf{Llama}} & \multicolumn{4}{c}{\textbf{GPT-5.5}} \\
\cmidrule(lr){2-5}\cmidrule(lr){6-9}\cmidrule(lr){10-13}\cmidrule(lr){14-17}
& \textbf{Mod.} & \textbf{Ver.} & \textbf{Time} & \textbf{Edits} & \textbf{Mod.} & \textbf{Ver.} & \textbf{Time} & \textbf{Edits} & \textbf{Mod.} & \textbf{Ver.} & \textbf{Time} & \textbf{Edits} & \textbf{Mod.} & \textbf{Ver.} & \textbf{Time} & \textbf{Edits} \\
\midrule
Example & \checkmark & \checkmark & 103.0 & 15 & $\times$ & $\times$ & 77.8 & 16 & $\times$ & $\times$ & 428.4 & 15 & \checkmark & \checkmark & 155.9 & 0 \\
NSPK & \checkmark & $\times$ & 2512.1 & 80 & $\times$ & $\times$ & 91.0 & 81 & $\times$ & $\times$ & 1048.3 & 93 & \checkmark & \checkmark & 2008.4 & 0 \\
Naxos & \checkmark & \checkmark & 264.7 & 47 & $\times$ & $\times$ & 52.6 & 43 & $\times$ & $\times$ & 435.5 & 49 & \checkmark & \checkmark & 958.8 & 0 \\
Toy & \checkmark & \checkmark & 40.5 & 11 & \checkmark & \checkmark & 29.2 & 40 & $\times$ & $\times$ & 1559.9 & 28 & \checkmark & \checkmark & 164.0 & 0 \\
Woo\&Lam & \checkmark & $\times$ & 4340.2 & 85 & $\times$ & $\times$ & 68.2 & 97 & $\times$ & $\times$ & 1272.4 & 97 & \checkmark & \checkmark & 965.47 & 0 \\
Sigfox & \checkmark & $\times$ & 1476.7 & 44 & $\times$ & $\times$ & 51.4 & 44 & $\times$ & $\times$ & 1546.2 & 50 & \checkmark & \checkmark & 241.5 & 0 \\
\midrule
X509.1 & \checkmark & \checkmark & 668.3 & 36 & $\times$ & $\times$ & 65.0 & 39 & $\times$ & $\times$ & 1024.2 & 39 & \checkmark & \checkmark & 414.3 & 1 \\
Denning & \checkmark & \checkmark & 5015.8 & 49 & $\times$ & $\times$ & 68.7 & 69 & $\times$ & $\times$ & 1356.0 & 69 & \checkmark & \checkmark & 2206.9 & 0 \\
KaoChow & \checkmark & \checkmark & 3888.8 & 75 & $\times$ & $\times$ & 202.8 & 102 & $\times$ & $\times$ & 2401.1 & 104 & \checkmark & $\times$ & 4484.61 & 0 \\
NSSK & \checkmark & \checkmark & 1588.9 & 74 & $\times$ & $\times$ & 100.3 & 97 & $\times$ & $\times$ & 2647.2 & 97 & \checkmark & \checkmark & 2286.0 & 0 \\
Stubblebine & \checkmark & $\times$ & 4201.0 & 115 & $\times$ & $\times$ & 97.4 & 114 & $\times$ & $\times$ & 601.1 & 118 & \checkmark & $\times$ & 5482.6 & 0 \\
OtwayRees & \checkmark & $\times$ & 5125.7 & 9 & $\times$ & $\times$ & 112.6 & 102 & $\times$ & $\times$ & 2321.1 & 111 & \checkmark & $\times$ & 6652.18 & 0 \\
Yahalom & \checkmark & $\times$ & 1468.3 & 93 & $\times$ & $\times$ & 154.4 & 101 & $\times$ & $\times$ & 1798.0 & 102 & \checkmark & $\times$ & 4014.83 & 0 \\
\midrule
EDHOC & \checkmark & $\times$ & 5161.6 & 102 & $\times$ & $\times$ & 72.6 & 130 & $\times$ & $\times$ & 1462.2 & 126 & \checkmark & \checkmark & 2360.2 & 0 \\
KEMTLS & \checkmark & \checkmark & 3771.8 & 88 & $\times$ & $\times$ & 67.9 & 127 & $\times$ & $\times$ & 543.5 & 128 & \checkmark & \checkmark & 2065.8 & 11 \\
LAKE & \checkmark & \checkmark & 3262.5 & 67 & $\times$ & $\times$ & 63.5 & 79 & $\times$ & $\times$ & 1598.6 & 78 & \checkmark & \checkmark & 2125.2 & 0 \\
SPLICE & \checkmark & $\times$ & 5606.9 & 182 & $\times$ & $\times$ & 102.9 & 187 & $\times$ & $\times$ & 1949.3 & 187 & \checkmark & \checkmark & 802.5 & 0 \\
SSH & \checkmark & $\times$ & 5401.4 & 130 & $\times$ & $\times$ & 71.4 & 125 & $\times$ & $\times$ & 1133.7 & 130 & \checkmark & \checkmark & 2287.0 & 0 \\
\midrule
\textbf{Total} & 18/18 & 9/18 & 53898.2 & 1302 & 1/18 & 1/18 & 1549.7 & 1593 & 0/18 & 0/18 & 25126.7 & 1621 & 18/18 & 14/18 & 39676.2 & 12 \\
\bottomrule
\end{tabular}
\end{table*}

\subsection{RQ3: Generated versus Manual Models}
\label{subsec:rq3}

We compare the 18 GPT-5.5-generated SAPIC+ models with the corresponding manual models from prior work ~\cite{mao2025llm}. This comparison is performed at the SAPIC+ level to avoid mixing modeling choices with artifacts introduced by compiler expansion into multiset-rewriting models.

\begin{table}[t]
\centering
\caption{Quantitative comparison between 18 GPT-5.5-generated and manual SAPIC+ models.}
\label{tab:model-detail-alignment}
\small
\setlength{\tabcolsep}{4pt}
\begin{tabular}{lrrr}
\toprule
\textbf{Modeling dimension} & \textbf{Precision} & \textbf{Recall} & \textbf{F1} \\
\midrule
Value provenance and setup       & 0.824 & 0.682 & 0.714 \\
Message structure            & 0.555 & 0.586 & 0.560 \\
Cryptographic dependencies   & 0.652 & 0.704 & 0.656 \\
Verification checks          & 0.516 & 0.488 & 0.478 \\
\bottomrule
\end{tabular}
\end{table}

\begin{table*}[t]
\centering
\caption{Per-protocol F1 scores for four modeling dimensions. Each
entry is the F1 of the generated model against the manual SAPIC+ reference.}
\label{tab:per-protocol-detail-f1}
\scriptsize
\setlength{\tabcolsep}{3.5pt}
\begin{tabular}{lrrrr}
\toprule
\textbf{Protocol} & \textbf{Provenance} & \textbf{Message} & \textbf{Crypto} & \textbf{Checks} \\
\midrule
CCITT-X509          & 0.62 & 0.39 & 0.79 & 0.56 \\
Denning--Sacco      & 0.71 & 0.60 & 1.00 & 0.33 \\
EDHOC               & 0.47 & 0.30 & 0.18 & 0.50 \\
Example             & 0.67 & 1.00 & 0.67 & 0.57 \\
KEMTLS              & 0.50 & 0.42 & 0.25 & 0.38 \\
Kao--Chow           & 0.71 & 0.51 & 0.75 & 0.39 \\
LAK06               & 0.75 & 0.30 & 0.36 & 0.22 \\
NSPK                & 0.78 & 0.73 & 1.00 & 0.48 \\
NSSK                & 0.71 & 0.51 & 0.68 & 0.45 \\
Naxos               & 0.67 & 0.47 & 0.62 & 1.00 \\
Neuman--Stubblebine & 0.71 & 0.40 & 0.84 & 0.16 \\
Otway--Rees         & 0.71 & 0.53 & 0.33 & 0.28 \\
SPLICE              & 0.70 & 0.38 & 0.62 & 0.41 \\
SSH                 & 0.62 & 0.42 & 0.18 & 0.36 \\
Toy                 & 1.00 & 0.83 & 1.00 & 1.00 \\
Woo--Lam            & 0.80 & 0.68 & 0.69 & 0.48 \\
Yahalom             & 0.71 & 0.65 & 0.94 & 0.41 \\
Sigfox              & 1.00 & 0.96 & 0.90 & 0.62 \\
\bottomrule
\end{tabular}
\end{table*}

Table~\ref{tab:model-detail-alignment} shows that setup and value provenance are the best-aligned dimension (F1 $=0.714$), followed by cryptographic dependencies (F1 $=0.656$). Alignment is weaker for message structure (F1 $=0.560$) and verification checks (F1 $=0.478$). Thus, generated models more often retain the protocol's cryptographic vocabulary than the exact conditions under which a received value may be trusted.

We selected representative cases from Table~\ref{tab:per-protocol-detail-f1} to illustrate three distinct outcomes: (i) close agreement on the protocol body under different threat or session assumptions, (ii) over-modeling reflected in low precision, and (iii) causal errors that are not captured by a bag of structural atoms.

(i) NSPK and Example show that agreement on the protocol body can coexist with differences in threat and session modeling. For NSPK, cryptographic-dependency F1 is $1.00$ and message-structure F1 is $0.73$, reflecting the preservation of the three encrypted messages. The manual model includes replicated honest principals and an explicit compromise interface. However, the generated model publishes one attacker-controlled private key and does not reproduce the same replicated-principal structure. Example makes this distinction even sharper. Its message-structure F1 is $1.00$, showing that client request, server decryption, and hash response have the same symbolic shape in both models. But its provenance F1 drops to $0.67$ because generated model omits the replicated client/server sessions and explicit long-term-key reveal branches present in the manual model. These differences matter because the same protocol body can be analyzed under different assumptions about principals, compromise, and session replication.

(ii) LAK06 and Woo--Lam illustrate how precision and recall expose different kinds of disagreement. LAK06 achieves provenance F1 of $0.75$, but its message-structure and cryptographic-dependency precision are only $0.30$ and $0.36$. The generated model introduces separate reader and backend branches, a prior-transcript role, additional forwarding steps, and duplicated outputs, while the manual model uses a compact synchronized-key protocol. The issue is therefore not omission but addition. Woo--Lam falls between the closely aligned and heavily over-modeled cases, with provenance, message, crypto, and checks F1 values of $0.80$, $0.68$, $0.69$, and $0.48$. It retains the broad message and cryptographic structure but differs more substantially in validation logic and adversarial events.

(iii) EDHOC and SSH expose a different problem. EDHOC has message-structure and cryptographic-dependency F1 values of $0.300$ and $0.18$, but the more important issue is that a single \texttt{HonestEDHOC} process constructs both initiator and responder messages sequentially. After sending the first message, it constructs the response locally and verifies its own locally generated signature without an intervening network input. The manual model instead separates initiator and responder processes with explicit \texttt{out}/\texttt{in} boundaries. SSH shows the same causal problem. Its cryptographic-dependency F1 is only $0.18$, but the more important issue is that the generated model does not preserve the distinction between locally generated and peer-provided values. Server values created locally are later treated as if they had been received from the peer. The model also omits the encrypted user-authentication request, acknowledgement, signed response, and final server verification present in the manual model.

Overall, the GPT-5.5-generated models match the manual references more closely on the protocol body than on security assumptions. Therefore, generated models are useful as semantic drafts, but not as direct replacements for manually developed models. Human review is still needed to check assumptions about the attacker, session state, role ownership, and when inputs are accepted. In several cases, the generated model simplified or omitted details that affect which traces are reachable. As a result, a model may verify successfully while still representing a weaker or different threat model from the manual reference.

\subsection{RQ4: Cost Analysis}
\label{subsec:rq4}
RQ4 evaluates the cost of introducing Protocol IR and confidence-guided review. Compared with direct LLM-to-SAPIC+ generation, our framework adds an explicit semantic checkpoint that makes modeling decisions inspectable but requires additional human effort. We evaluate workflow cost in terms of cell-level edits, human inspection and repair effort, and model generation overhead. The "Edits" column in Table~\ref{tab:e2e} counts UI-level corrections to the generated IR before SAPIC+ generation, including fixes to message fields, value provenance, role-local knowledge, verification targets, and compromise assumptions. DeepSeek, GPT-4o, and Llama require 72.3, 88.5, and 90.1 edits
per protocol on average, respectively, while GPT-5.5 averages only 0.7 edits and reaches highest proof verification success. Smaller protocols such as Example, Toy, and Otway Rees require relatively few edits, whereas larger and more complex protocols such as KEMTLS, SPLICE, and SSH require substantially more corrections. This suggests that model ability and protocol complexity directly affect the amount of human repair needed at the IR level.

We also report model generation and verification time in
Table~\ref{tab:e2e}. We also report model generation and verification time in Table 2. These measurements record the observed end-to-end wall-clock time for translating the reviewed IR into SAPIC+ and running Tamarin. However, the times are outcome-conditioned. Failed runs may terminate before proof search, whereas accepted models may go through the full verification process. The totals should therefore not be used to directly compare efficiency across models. Together with the manual edit counts, these results characterize the main trade-off of our framework: the Protocol IR introduces an additional review step, but this step makes the modeling process more transparent, repairable, and trustworthy.

%% file: 2case.tex
\section{Case Study}
\label{subsec:case-study}

We further present a case study on \texttt{KEMTLS}, a protocol whose natural-language description is difficult to translate directly into a formal model because several protocol steps are described as phases and flights rather than as individual messages. In particular, the input states:

  \begin{quote}
  \emph{"In the same server-to-client flight as ServerHello, the server also sends a certificate containing its long-term KEM public key."}
  \end{quote}

\begin{figure}[h]
  \centering
\includegraphics[width=\linewidth]{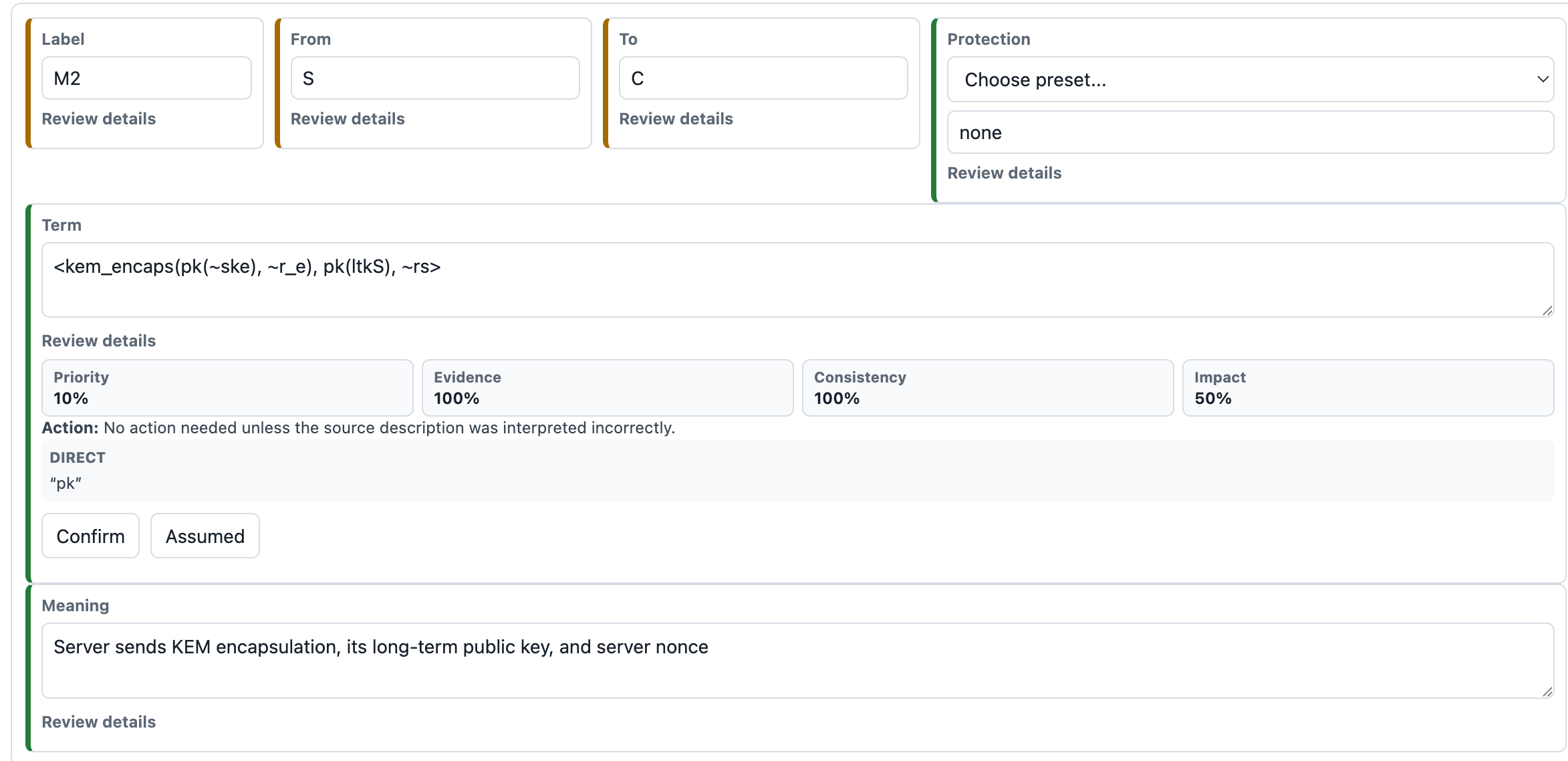}
    \label{fig:kemtls-raw-m2}
  \caption{Raw IR fields for the partial KEMTLS review example. The raw IR represents server flight as a single plaintext message row \texttt{M2}.}
  \label{fig:kemtls-ui-raw}
\end{figure}

This sentence is easy for an LLM to misinterpret. The initial LLM-generated IR treated the server's long-term KEM public key as if it were directly included in the unprotected \texttt{ServerHello} message. As shown in Figure~\ref{fig:kemtls-ui-raw}, the raw IR models the server response as a single plaintext \texttt{M2} row, $\langle \mathsf{kem\_encaps}(\mathsf{pk}(\tilde{ske}), \tilde r_e),
\allowbreak\ \mathsf{pk}(ltkS),\allowbreak\ \tilde{rs}\rangle$, thereby placing the KEM encapsulation, the server long-term public key, and the server nonce in the same unprotected message. This loses an important semantic distinction because \texttt{ServerHello} establishes the first shared secret, whereas the server certificate is a later handshake item protected by the derived server handshake traffic secret.
The review UI makes this error visible at the IR level, but not by simply marking the raw \texttt{M2} term as unsupported. The cells of this row, such as \texttt{ServerHello}, the KEM encapsulation, the server public key, and the nonce, are all mentioned in the source text. Thus, the \texttt{Term} and \texttt{Meaning} cells receive strong evidence confidence. The ambiguity instead appears in how those cells are assigned to protocol steps. In the raw IR, the "Checks" field is marked as high priority: the "role" and "source message" cells for the early KEMTLS checks are "must review", with review priority 100\%, evidence confidence 0\%, and semantic impact 100\%. 
During review, the user can repair this by splitting the server flight into separate steps. The reviewed excerpt separates \texttt{M2}, a plaintext \texttt{ServerHello}, from \texttt{M3}, an encrypted \texttt{ServerCert} protected under \texttt{SHTS}. The reviewed IR therefore models $M2 = \langle \texttt{SERVERHello}, cte, rs \rangle$ and then $M3 = \mathsf{senc}(\langle \texttt{ServerCert}, cert(pkS) \rangle, SHTS)$.

This repair also affects the later checks. The reviewed IR explicitly records that the client first derives \texttt{CHTS} and \texttt{SHTS} from \texttt{ServerHello}, then decrypts \texttt{ServerCert}, recovers the server KEM public key, and only then sends the client KEM ciphertext and finished messages. These UI-visible edits prevent SAPIC+ generator from proving properties over a model in which the server authentication key is available at the wrong protocol point.
After review, the generated SAPIC+ model compiles cleanly and Tamarin matches all five reviewed KEMTLS target outcomes. This case illustrates why the IR layer is necessary. Tamarin can verify a formal model against formal lemmas, but it cannot decide whether a phrase such as "same flight" has been translated into the needed message structure. The IR therefore acts as a human-auditable checkpoint between natural language and formal verification.

%% file: 5discussion.tex
\section{Discussion}
\label{sec:discussion}

\subsection{Non-standard Cases}
\label{subsec:nonstandard-cases}

To examine whether the workflow remains usable beyond textbook protocols, we
applied it to three additional security scenarios: MCP session hijacking, GitHub Actions artifact provenance, and AWS external-ID delegation~\footnote{The implementation, reviewed inputs, generated models, and proof summaries are available at \url{https://github.com/laplace1002/TamarinAgent}.} Because unlike benchmark protocols, these cases are described mainly through platform documentation, attacks, and mitigations rather than complete message sequences.

\lstdefinestyle{casecode}{
  basicstyle=\ttfamily\scriptsize,
  keywordstyle=\color{blue!70!black},
  commentstyle=\itshape\color{black!55},
  stringstyle=\color{orange!70!black},
  backgroundcolor=\color{white},
  rulecolor=\color{black},
  frame=tb,
  framerule=0.55pt,
  framesep=3pt,
  xleftmargin=0pt,
  framexleftmargin=0pt,
  numbers=none,
  columns=fullflexible,
  keepspaces=true,
  breaklines=true,
  breakatwhitespace=false,
  showstringspaces=false,
  aboveskip=1pt,
  belowskip=1pt,
  morekeywords={rule,restriction,lemma,exists-trace,all-traces},
  morestring=[b]',
  morecomment=[l]{//},
  morecomment=[s]{/*}{*/}
}

\paragraph{MCP session hijacking.}
The MCP case shows how the workflow converts a real attack scenario into a model that can be checked automatically. The model captures both session-hijacking methods described in the source, including injecting messages into a shared session queue and directly impersonating another session. Tamarin reproduces both attacks and shows why they are possible. A request is accepted when its session identifier matches, without first checking whether the sender is authorized to use that session.

\begin{figure}[H]
\begin{lstlisting}[style=casecode]
rule MCP_Receive:
 [ St_MCP(sid), In(call(sid_call, request)) ]
 -->
 [ St_MCP_Check(sid, sid_call, request) ]

rule MCP_CheckSession:
 [ St_MCP_Check(sid, sid_call, request) ]
 --[ Pred_Eq(sid_call, sid) ]->
 [ St_MCP_Accept(sid, request) ]

rule MCP_Accept:
 [ St_MCP_Accept(sid, request) ]
 --[ CallAcceptedAsSession(
       'MCPServer','Attacker','Client',sid,request) ]->
 [ ]
\end{lstlisting}
\caption{Tamarin MSR rules for MCP session acceptance.}
\label{fig:mcp-session-acceptance}
\end{figure}

Figure~\ref{fig:mcp-session-acceptance}~\footnote{For readability, we alpha-rename only compiler-generated state facts and omit state arguments that do not participate in the illustrated decision} shows the key modeling choice of this attack. The received session identifier is checked against the stored identifier, and the compiler-generated \texttt{predicate\_eq} gives this comparison its equality semantics. However, the authorization lemma requires an earlier \texttt{AuthorizedInboundRequest} event, which is absent from the acceptance path. The counterexample therefore shows that a valid session identifier is sufficient to associate a request with a session, but not to authenticate the sender.

The verification results also need to be read according to the type of property being checked. Tamarin verifies the two reachability lemmas because it finds the expected attack traces, while the authorization lemma is falsified by a counterexample. Thus, the model captures both the intended attack paths and the missing authorization check.

\paragraph{GitHub Actions artifact provenance.}
The GitHub Actions case compares three artifact-selection policies. Selecting an artifact only by name allows an artifact from another run to be substituted. Binding the artifact to the triggering run prevents this cross-run substitution, but it does not guarantee that the run itself came from a trusted source because a fork run can still be the triggering run. The stronger \texttt{trusted\_origin} policy therefore also checks repository and branch metadata.

\begin{figure}[H]
\begin{lstlisting}[style=casecode]
rule RunBound_Select:
 [ St_RunBound(producing_run, trigger_run,
               artifact, repo, branch) ]
 --[ Pred_Eq(producing_run, trigger_run),
     ArtifactSelected('run_bound', trigger_run,
                      artifact, trigger_run, repo, branch) ]->
 [ ]

rule TrustedOrigin_Select:
 [ St_Trusted(producing_run, trigger_run,
              artifact, repo, branch) ]
 --[ Pred_Eq(producing_run, trigger_run),
     Pred_Eq(repo, 'protected_repo'),
     Pred_Eq(branch, 'protected_branch'),
     ArtifactSelected('trusted_origin', trigger_run,
                      artifact, trigger_run,
                      'protected_repo','protected_branch') ]->
 [ ]
\end{lstlisting}
\caption{Tamarin MSR rules for run-bound and trusted-origin artifact selection.}
\label{fig:github-policy-difference}
\end{figure}

Figure~\ref{fig:github-policy-difference} highlights this distinction. The run identifier establishes which execution produced the artifact, while the repository and branch checks establish whether that execution came from an allowed source. As a result, \texttt{run\_bound} satisfies the run-provenance property but still admits a fork trace, whereas \texttt{trusted\_origin} blocks that trace. An executability lemma also confirms that the stronger policy still allows a legitimate protected-branch deployment, rather than achieving security by rejecting all deployments.

Our workflow automatically identifies nine verification targets covering three questions. Theses are whether a fork artifact can reach privileged use, whether the selected artifact belongs to the triggering run, and whether that run has a trusted origin. All nine results match the expected outcomes. Assuming that the metadata used for these checks is authentic, the case shows that binding an artifact to the correct producer run is not sufficient to establish that the run originated from an authorized source.

These non-standard cases highlight a different strength of the workflow from the benchmark evaluation. Even without a complete reference model, it can turn security assumptions described in natural language into explicit policy variants and verification targets. This makes differences such as session identity versus sender authorization, and artifact provenance versus trusted origin directly testable. By checking attack reachability, safety, and legitimate execution separately, the workflow also makes clear why a policy succeeds or fails rather than reducing the result to a single verification outcome.

\subsection{Limitations and Future Work}
\label{subsec:discussion-limitations}

Our approach does not fully solve the problem of interpreting natural-language protocol
descriptions. If the source description omits important details, the system may still require
the user to supply missing assumptions. Moreover, the quality of the generated IR depends on
the ability of the LLM to identify relevant protocol concepts and provide useful evidence.
The confidence mechanism can help expose uncertainty, but it cannot guarantee that all semantic
errors will be found. Finally, the current framework focuses on improving the modeling workflow
rather than proving the correctness of the IR-to-SAPIC+ translator. A stronger implementation
would require a more formal definition of the IR semantics and a verified or systematically
validated translation procedure.

As future work, we plan to add a lightweight review step that compares the generated model with the original description. LLM would check whether the model captures the messages, checks, events, and compromise assumptions stated in the description, and whether it introduces any behavior or assumptions that the description does not support. It would also flag missing steps or changes in their order. This could help users identify potential modeling errors.

%% file: 7conclusion.tex
\section{Conclusion}
\label{sec:conclusion}
This paper presented a confidence-guided Protocol IR for LLM-aided security protocol modeling. 
By introducing an explicit semantic checkpoint between natural-language interpretation and formal model generation, our approach makes critical modeling decisions inspectable, correctable, and auditable before formal verification. 
The confidence-guided interface further helps users focus on uncertain and high-risk fields. 
Overall, the proposed trust boundary provides a practical foundation for trustworthy LLM-assisted security protocol verification.